# Quantum Phase Transition Induced by a Preformed Pair in a Boson-Fermion Model of Fulleride Superconductivity

Richard H. Squire$^{\xi}$, Norman H. March$^{*\dagger}$

$^{\xi}$ Department of Chemistry, West Virginia University
Institute of Technology University
Montgomery, WV 25303, USA

$^{*}$ Department of Physics, University of Antwerp (RUCA), Groenborgerlaan 171,
B-2020 Antwerp, Belgium
and
Oxford University, Oxford, England

There continues to be enormous interest in the BCS to BEC transition. While the BCS and BEC "end points" seem to be well-established, in the intermediate region – home to fulleride and high temperature copper oxide superconductors – considerable extrapolation of the models must be done as there still is no exact theory. We recently reported a revealing reinterpretation of the condensed phase Boson-Fermion Model (BFM) by comparing it to a "cold" atom formulation [1]. While the ground and singly excited states appear to remain continuous in all models we have examined, the collective modes contain a singularity due to a Feshbach resonance (tuned by doping) causing a breakdown of the Migdal theorem. As a result of vertex corrections, there is a fundamental change in the nature of the superconductivity due to the formation of "preformed pairs" as the previously suggested location [1] of a quantum critical point in the fulleride phase diagram is passed. The result is a quantum phase transition (QPT) between BCS and BEC-like (or Feshbach resonance) superconductivity (SC).

We discuss features of the resonance and the role of the experimentally observed preformed pair formation in fullerides, essential to the Boson-Fermion Model (BFM), and often speculated since the work of Nozieres and Schmitt-Rink [17]. Here, we present arguments to establish a model of the preformed pair which can be favorably compared to a circular charge density wave (CDW) isolated on a fulleride molecule. The binding is much larger than a Cooper pair. The CDW seems to be stabalized by splitting of the Jahn-Teller active $H_g$ vibrational modes to reduce Coulomb repulsions. Our conclusions are: 1) the doping of two electrons into triply degenerate $t_{1u}$ orbitals results in the experimentally observed singlet state (CDW); and 2) this CDW (preformed pair) has a dual role as doping is varied: suppression of BCS SC and enabling a Feshbach resonance form of SC.

**1. Introduction.**

The study of fulleride and high temperature superconductivity (HTSC), principally the copper oxide systems continues with high intensity. Our previous work suggested that the two forms of superconductivity (SC) in both of these systems are a result of the level of doping which changes the nature of the interaction from a Feshbach resonance at low doping to a BCS-type superconductivity at higher levels [1]. To construct a model which extrapolates from the BEC region to the BCS regime we have used a modified Boson-Fermion model (BFM), discussed previously [2]. To briefly recap this work, the highly doped side of the fulleride superconductivity (SC) "dome" begins with a doping level of 4 electrons into an insulating state, and ends with two doped electrons (Fig 1) creating another insulating state. On the highly doped side, the SC is BCS-like. As the doping level is reduced (somewhere in the region between 4 and 3 electrons), a "pre-formed" electron pair (PP) is formed, localized on a fulleride molecule. In the phase diagram the PP dissociation or ionizing line emerges near the top of the dome and continues to higher energy. We are suggesting that this PP creates the pseudogap discussed widely in the literature [3]. In this intermediate region, roughly where the free energy $\mu = 0$, a demarcation point between BCS $(\mu > 0)$ and BEC $(\mu < 0)$ and home to fulleride and high temperature superconductors, there is not only a pre-formed pair but also a quantum critical point created by a singularity in the two body scattering length due to the Feshbach resonance as we discussed earlier [1]. As discussed, even though the two body scattering length changes discontinuously at the unitary scattering condition, superconductivity still varies smoothly as does the free energy and the one body excitation energy. However, there is a fundamental change in the nature of the superconductivity when weakly-bound BCS Cooper pairs become dominated by the strongly bound pre-formed pair as the previously suggested location of a quantum critical point in the fulleride phase diagram is passed. Despite the lack of experimental evidence, one can estimate where the QCP should approximately be since it is at unitarity scattering. One of the features of a QCP is presumably its far reaching influence which is clearly present in the property changes that occur as a fulleride (or high-$T_c$ SC) is passed from one side of the superconducting "dome" to the other, which we discuss in detail.

The outline of this manuscript is as follows: Section 2 is a brief discussion of the origin of the BFM. Section 3 contains for perspective Cooper's argument for the formation of weak electron pairs along with the significantly different preformed pair, localized on a single fulleride molecule. Section 4 discusses the preformed pair as a localized charge density wave (CDW), stabilized by certain Jahn-Teller $H_g$ molecular vibrations along with general experimental and theoretical support, followed by a CDW mean field theory. Section 5 discusses previous work concerning SC and CDW interactions. Also discussed is the form of a possible wave function for a fulleride molecule along with some possible new properties. Section 5 contains information relevant to SC and CDW interactions and preformed pairs, followed by other pertinent work concerning preformed pairs in Section 6. Section 7 illustrates the general theory of a preformed pair can be included in the BFM by us and others. Section 8 illustrates the general theory along with a specific application to fullerides. Conclusions and suggestions for future work follow in Section 9.

## 2. Why Use the Boson-Fermion Model (BFM) at All?

a. Development of the BFM. The BFM was created by Ranninger and collaborators [4] who began work on the model in a different context where the bosons were considered as phonons (more precisely, polarons) instead of electron pairs. Considerable exploration of the model followed including a proposal for the existence of a pseudo gap phase [4b], mean field solutions, a "restart" [5], concluding later with resonating polarons [6] which has some similarities to our work.

Shortly after the discovery of HTSC Friedberg and Lee [7] adopted the BF model based on a speculated resonance in particle work. Motivated by the short coherence length, and previous work by Schafroth [8], who prior to the BCS theory had proposed an equilibrium between electrons and a paired complex, these authors argued for real space as opposed to momentum space pairing, as is in the BCS model. Interestingly enough, one of the conclusions of this early work (almost 20 years ago) suggests the existence of a superfluid for fermions as well as bosons.

b. Our use of the Model. Our interest in the BFM arose from a comparison of it with the Su-Schrieffer-Heegler (SSH) model of polyacetylene (PA) [9] applied to fullerides [10]. Initially the phonons of SSH appeared to have a similar influence such that fullerides could be attributed to a "classical" BCS phonon-induced SC mechanism with possible topological character, favorably compared with a modification of the BFM [11]. When we became aware that certain similarities of the $t_{1u}$ plasmon band (0.5 eV with a width of 0.5 eV) seemed to indicate a resonance similar to the Kondo problem, we embarked on a review of the Anderson-Kondo problem [12]. Included in a general form of the BFM is a Kondo-like condition [13]. In addition a fulleride molecule doped with two electrons seems to have Wigner-like crystal correlations [14], again leading us to the BFM. Finally, cold atom theorists have used the BFM to describe their work with the BCS term of the Hamiltonian "turned on" [1, 15] which results in the BCS-preformed pair interaction, which we recognized as appropriate to fulleride and HTSC. Thus, a quantum phase transition is permitted (see Section 7 for details).
.
## 3. Cooper and Preformed Electron Pairs

### a. Cooper's pair

The origin of electron pairing in a superconductor was successfully studied by Cooper [16] using a velocity-dependent potential V for two electrons interacting above a noninteracting (filled) Fermi sea. The Fermi sea marks the boundary between filled energy levels and those which are empty; the only contribution of the filled sea is to deny entry into Fermi sea states due to the Pauli principle. States above the Fermi sea will then have kinetic energy $\varepsilon_k > 0$. We will assume that the total spin and center of mass momentum $\hbar \vec{q}$ of the pair are constant, so an orbital wave function of the pair is

$$\psi(\vec{r}_1, \vec{r}_2) = \varphi_q(\vec{\rho}) e^{i\vec{q} \cdot \vec{R}} \qquad (1)$$

with center of mass and relative coordinates defined as $\vec{R} = (\vec{r}_1 + \vec{r}_2)/2$ and $\vec{\rho} = \vec{r}_1 - \vec{r}_2$, respectively. As $\vec{q} \to 0$ the relative coordinate is spherically symmetric and hence, $\varphi(\vec{\rho})$ is an eigenfunction of angular momentum with angular momentum quantum numbers l and m. If $\vec{q} \neq 0$, l is not a good quantum number, but the component of angular momentum along $\vec{q}$ and parity are.

Assuming $\vec{q} = 0$, $\psi$ can be expanded as

$$\psi(\vec{r}_1, \vec{r}_2) = \varphi_q(\vec{\rho}) = \sum_k a_k e^{i\vec{k} \cdot \vec{r}_1} e^{-i\vec{k} \cdot \vec{r}_2} \qquad (2)$$

where the sum is restricted to states with $\varepsilon_k > 0$. If $e^{i\vec{k} \cdot \vec{r}_1}$ and $e^{-i\vec{k} \cdot \vec{r}_2}$ are thought of as plane wave states, then the pair wave function is a superposition of definite pairs where $\pm \vec{k}$ is occupied. We can write a Schrödinger equation

$$(H_0 + V)\psi = E\psi \quad \text{or} \quad (E - 2\varepsilon_k) a_k = \sum_{k'} V_{\vec{k}\vec{k}'} a_{\vec{k}'} \qquad (3)$$

with matrix element $V_{\vec{k}\vec{k}'}$ defined as

$$V_{\vec{k}\vec{k}'} = \langle \vec{k}, -\vec{k} | V | \vec{k}', -\vec{k}' \rangle \qquad (4)$$

There is no general solution to Eq (3) unless $V_{\vec{k}\vec{k}'}$ is assumed to be factorizable,

$$V_{\vec{k}\vec{k}'} = \lambda w_{k'}^* w_k \qquad (5)$$

as was done by Nozieres and Schmitt-Rink [16] in their BCS to BEC crossover study. In an isotropic system $V_{\vec{k}\vec{k}'}$ can be expanded into partial wave components

$$V_{\vec{k}\vec{k}'} = \sum_{l=0}^{\infty} \sum_{m=-l}^{l} V_l(|\vec{k}|, |\vec{k}'|) Y_l^m(\Omega_{\vec{k}}) Y_l^{-m}(\Omega_{\vec{k}'}) \qquad (6)$$

and the pair (l,m) eigenstates can be determined

$$V_l(|\vec{k}|, |\vec{k}'|) = \lambda_l w_k^l w_{k'}^{l*}$$

Then,

$$(E - 2\varepsilon_k) a_k = \lambda_l w_k^{l*} \sum_{k'} w_{k'}^{l*} a_{k'} \qquad (7)$$

with

$$a_{\vec{k}} = a_k Y_l^m(\Omega_{\vec{k}})$$

Rearranging eq (7)

$$a_k = \frac{\lambda_l w_k^l C}{E_{lm} - 2\varepsilon_k} \qquad (8)$$

with
$$C = \sum_k w_k^{l*} a_k \quad (9)$$

substituting eq (8) into (3) then yields,

$$1 = \lambda_l \sum_k |w_k^l|^2 \frac{1}{E_{lm} - 2\varepsilon_k} \equiv \lambda_l \Phi(W_{lm}) \quad (10)$$

If we assume we are working in a large, but finite box, we can graphically display the single particle energies $\varepsilon_k$ which form a discrete set such that when $W_{lm}$ passes from below to above $2\varepsilon_k$, $\Phi(W)$ jumps from $-\infty$ to $+\infty$ as W repeats this passage for multiple levels (Fig 2 illustrates 10 levels). Thus, the eigenvalues $W_{lm}$ for attractive (negative) $\lambda_l$, are shown as $1/\lambda_l < 0$. The key feature Cooper discovered is that a bound state is split out of the continuum for an attractive potential, no matter how weak. The binding energy of the pair, $|W_{lm}|$, which is split-off is

$$\frac{1}{|\lambda_l|} = \frac{N(0)}{2} \log\left[\frac{|W_{lm}| + 2\omega_c}{|W_{lm}|}\right]$$

which, if we assume that the slowly varying density of states can be approximated by $N(0)$, the density of single-electron states of a single spin orientation at the Fermi surface, becomes

$$|W_{lm}| = \frac{2\omega_c}{\exp\left[\frac{2}{N(0)|\lambda_l|}\right] - 1}$$

At weak coupling where $\left[N(0)|\lambda_l| << 1\right]$

$$|W_{lm}| \simeq 2\omega_c \exp\left[-\frac{2}{N(0)|\lambda_l|}\right] \quad (11)$$

Since these pairs are weakly coupled, about $10^6$ pairs overlap in a typical superconductor (see Fig 3 for general pair size plotted vs. the product $k_F \xi$ as a natural variable and references in [1] for a general discussion).

**b. Plausibility of a preformed pair**

A preformed pair needs to be distinguished from a Cooper pair (CP) of BCS theory which almost immediately after formation, condenses into a superconductor. A CP is a weakly bound pair of electrons held together by an attraction due to exchange of virtual phonons; it does not obey boson commutation rules. It is quite large by molecular standards (which is why it is usually represented in momentum space), as it covers some $10^{-4} cm^2$ in area where it overlaps with a large number of other CP's. Nonetheless, the wave functions of all these CP's are orthogonal and correlated. A Bose-Einstein (BE) pair on the other hand is a strongly bound pair of electrons that is a true boson. It forms at a temperature, $T^*$, much above $T_c$, the superconducting temperature, has a small pair size (real space pairing) and can be thought of as an ideal gas. The parameters for a preformed pair are somewhere in between these two limiting cases.

First, we want to make a general plausibility argument for a preformed pair. Use of a band model suggests fullerides with two doped electrons would be metallic, not an insulator as experimentally found. The standard molecular orbital model suggests that the lowest unoccupied energy levels contain three degenerate $t_{1u}$ orbitals; two doped electrons would produce a triplet if one of Hund's rules is followed. Clearly something unusual is happening. Prior to Cooper's work, Fröhlich [18] predicted that despite the repulsive interaction from Coulomb effects, two electrons near the Fermi surface can attractively interact with each other (see Section 4a for his Hamiltonian and more discussion). We can make further progress if we use a variant of the March model [19] as applied to fullerides by Clougherty [20]; the two electrons on a fulleride "sphere" will most likely be on opposite sides of the sphere due to electron-electron repulsion (strictly speaking C60 is not a sphere, but an icosahedron with vertices on a sphere since it is a Platonic solid).

The result of doping of two electrons onto a fulleride molecule produces a Jahn-Teller (JT) effect. The result is an orbital triplet, $t_{1u}$, with both electron residing in the same orbit producing a singlet state, violating Hund's rule and with an additional important consequence that this state is insulating. Somehow these electrons are being contained on a fulleride molecule. Thus, a fulleride molecule needs effective screening of Coulomb repulsions to be able to contain multiple electrons. But we anticipate this repulsion to be significantly lower than bare electrons due to screening by other electrons and vibrations of the fulleride molecule. Calculations have produced a Hubbard U value of 3.0 eV [21]. In the general case, since hopping is greatly reduced, screening would also reduce the electron-phonon interaction, but since a fulleride molecule is a molecular solid with multiple vibrational modes, these modes can interact very differently based on the number of intermolecular phonons and their symmetry [21]. The comparison of the JT active $A_g$ and $H_g$ modes in the presence of an increased Coulomb charge, demonstrates how different symmetries respond. The energy of the $t_{1u}$ level for an $A_g$ mode as a function of a phonon coordinate Q moves to higher energy because it is totally symmetric and cannot distort to better accommodate the charge. However, the same charge for an $H_g$ mode will force the mode to split into $D_{3d}$ and $D_{5d}$ reduced symmetries while maintaining a barycenter. This is similar to symmetry splittings in a crystal field as illustrated by d orbitals or Tanabe-Sugano diagrams. The resulting symmetries with "peaks" and "valleys" on the surface of the (almost) spherical fulleride surface lead to a ring (see Fig 2.7 and 2.8 [22] and the next section). The result is an efficient Coulomb screening while maintaining the effectiveness of the $H_g$ electron-phonon interaction in an anisotropic manner.

Examining the possible structures for the fulleride pre-formed CP using the Weisskopf CP model [23] where one electron moves in one direction and creates a "tube" of attraction for the second electron moving in the opposite direction, we envisage a distortion around the entire fulleride sphere that is similar to a "distorted great circle" via the Jahn-Teller effect (as previously discussed [11, 24, 25]) with the $t_{1u}$ state interacting with $H_g$ vibrations. We can argue that the distortion due to a minimization of the Coulomb potential becomes an anisotropic confining potential for the two doped electrons and an additional deformation potential is created by passage of an electron in the preformed pair, thereby creating a further increase in binding energy for the second electron. As a result, the binding is much stronger than Cooper's model. To more quantitatively describe our model for the PP, we return to Fröhlich's work below.

There are several suggestions as to how the singlet insulator might occur [26, 27, 28, 29]. We favor an alternative suggestion, namely that the spin triplet state is unstable with respect to the formation of a localized tightly bound electron pair (PP) above $T_c$ and below some $T^*$ (see Fig 1), such that the energy gained, $E_I$, (significantly stronger than a BCS CP) is sufficient to negate Hund's rule. Essentially, two spin-paired electrons are localized on a fulleride in a "special" circular charge density wave to our knowledge not previously described in the literature. We maintain that this entity, presumably experimentally established in fullerides, could be considered as analogous to the elusive "pre-formed pair", heretofore speculated about in high-$T_c$ SC studies (more on this later). The result is that the JT orbital triplet is split with the pair in the occupied orbital now lower in energy and assuming CDW character, we can then approximate what the stabilization energy might be (next section).

Assuming this is the case, the pseuodogap origin seems apparent; it is due to a localized pairing without long range order (LRO), as the phases of the wave functions of the PP on adjacent fulleride molecules are random. This proposal also suggests why the metal-insulator (M-I) transition that certain doped fullerides span happens; small changes in a fulleride structure can move seemingly closely related fullerides from one side of the transition to the other. The possibility that doping a third electron onto a fulleride can cause an energy change larger that energy $E_I$ would suggest that the metal would have three unpaired electrons. This possibility is inconsistent with our superconductivity theory, so we are left with the interesting possibility that the PP remains intact upon doping with a third electron. The resulting material might be perceived as a "poor conductor", as the possibility exists that the PP's remain localized above $T_c$. This material, say $Rb_3C_{60}^{3-}$, might be described as a "conductor with boson insulating features." It is interesting to note that with a modest change of energies in Gunnarsson's arguments [30], one can arrive at our conclusion even though that was not the original intent; this is a reflection of the subtlety of this particular M-I transition where it seems that structurally related fullerides must be determined case by case as to whether they are metals or insulators.

**c. How does a preformed pair exist in a Jahn-Teller environment?**

We have previously discussed a doped fulleride molecule in term of the Jahn-Teller effect [11, 31]. So what form does the preformed pair have in this environment? A brief review begins with the coupling of the eight, five dimensional $H_g$ vibrations with the three $t_{1u}$ orbitals, represented in the icosahedral group by the spherical harmonics $\{Y_{2m}\}_{M=-2}^{2}$ and $\{Y_{1m}\}_{m=-1}^{1}$, respectively. After creating a second quantized Hamiltonian in a real representation is

$$H = H^0 + H^{e-v}$$

$$H^0 = \frac{\hbar\omega}{2}\sum_\mu \left(-\partial_\mu^2 + q_\mu^2\right)$$

$$H^{e-v} = g\frac{\hbar\omega}{2}\sum_s \left(c_{xs}^\dagger, c_{ys}^\dagger, c_{zs}^\dagger\right) \begin{pmatrix} q_0+\sqrt{3}q_2 & -\sqrt{3}q_{-2} & \sqrt{3}q_1 \\ -\sqrt{3}q_{-2} & q_0-\sqrt{3}q_2 & -\sqrt{3}q_{-1} \\ -\sqrt{3}q_1 & \sqrt{3}q_{-1} & -2q_0 \end{pmatrix} \begin{pmatrix} c_{xs} \\ c_{ys} \\ c_{zs} \end{pmatrix}$$

This Hamiltonian is rotational invariant so simultaneous rotations of electronic and vibronic representations have no effect on the eigenvalues as O'Brien and Longuet-Higgins noted some time ago [31a, 32].

The distortions can be obtained on the classical limit by ignoring the vibron derivatives and treating $\vec{q} = \{q_\mu\}$ in $H^{e-v}$ as frozen and diagonalizing the coupling matrix (see [22, 31] for details.) Here, $\varpi = (\phi, \theta, \psi)$ are the Euler angles of the $O(3)$ rotation matrix T. Then, the electron energies depend on only two vibron coordinates,

$$\vec{q} = \begin{pmatrix} r \\ 0 \\ z \\ 0 \\ 0 \end{pmatrix}$$

The vibron coordinates $\vec{q}$ can be rotated to the diagonal basis using the L=2 rotation matrix $D^{(2)}$ whereby

$$\vec{q}_\mu = (r, z, \varpi) = \sum_{m,m',\mu'=-2} M_{\mu,m} D_{m,m'}^{(2)}(\varpi) M_{m',\mu'}^{-1} \vec{q}_{\mu'}(0)$$

($M_{\mu,m}$ is defined in [31a]). From this equation and the unitarity of D and M, $|\vec{q}|^2$ is invariant under rotations of $\varpi$. Consequently, the adiabatic potential energy V is dependent only on r, z, and the occupation numbers $n_i$, $\sum_i n_i = n$ of the electronic states.

$$V(z,r,[n_i]) = \hbar\omega/2 (z^2 + r^2) + \hbar\omega g/2 \left[n_1(z-\sqrt{3}r) + n_2(z+\sqrt{3}r) - n_3 2z\right]$$

The minimum in the classical energy is at the JT distortions $(\bar{z}_n, \bar{r}_n, \bar{n}_i)$ which are listed in Table I [29b] for different electron fillings. In the principle axes frame where $\tilde{\phi}, \tilde{\theta}$ are the longitude and latitude (coordinates 1 and 2), respectively and 3 is the north pole, the JT distortion in a real representation is

$$\left\langle u^{JT}\left(\tilde{\phi},\tilde{\theta}\right)\right\rangle = \frac{\bar{z}}{2}\left(3\cos^2\bar{\theta}-1\right)+\frac{\bar{r}\sqrt{3}}{2}\sin^2\bar{\theta}\cos\left(2\bar{\phi}\right) \qquad (12)$$

What has been accomplished is that the original five q's transform like a set of d states in the same space as the orbital triplet basis transforms as a set of p states. Then, if we hold $\bar{z}$ constant and $\bar{r}=0$, the representative point $\left(\bar{\phi},\bar{\theta}\right)$ moves on a two-dimensional surface on the surface of a sphere in three dimensional space. Our main interest is the n = 2 mode which has a uniaxial distortion about the 3 axis (see Fig 4). Quantum fluctuations about the frozen JT distortion are necessary in a semi-classical quantization. Five dimensional coordinates r, z, $\varpi$ can be defined, where $\varpi$ characterizes the "wine bittle" potential characteristic of the JT effect, while $r, z$ are transverse to the JT manifold. For finite distortions the kinetic energy can be shown to be

$$\frac{1}{2}|\dot{\vec{q}}|^2 = \frac{1}{2}\left[\dot{z}^2+\dot{r}^2+\sum_{l=1}^{3}I_i\omega_i^2\right]$$

which is the kinetic energy of a rigid body rotator with $I_i\left(\bar{z},\bar{r}\right)$ as the moments of inertia in the principle axis frame. For $\vec{r}\ll\bar{z}$ the potential energy is

$$V\left(\vec{r}\right)\approx \tfrac{1}{2}|\vec{r}|^2 \ ;$$

The potential is harmonic. Summarizing, the semi-classical unimodal distortion Hamiltonian is

$$H^{uni}\approx H^{rot}+H^{HO}$$
$$H^{rot}=\frac{\hbar\omega}{6\bar{z}^2}\vec{L}^2 \qquad (13)$$
$$H^{HO}=\hbar\omega\sum_{\gamma=1}^{3}\left(a_\gamma^\dagger a_\gamma+\tfrac{1}{2}\right)$$

with $\vec{L}$ an angular momentum operator and $H^{HO}$ described by three harmonic oscillator modes. The energies are, as expected from the discussion above,

$$E^{uni}=\hbar\omega\left\{\frac{1}{6\bar{z}_n^2}L(L+1)+\sum_{\gamma=1}^{3}\left(n_\gamma+\tfrac{1}{2}\right)\right\} \qquad (14)$$

### d. Agreement with experiment?

Experimental support for anisotropy receives support from the interpretation of the Raman spectra obtained by Winter and Kuzmany [33]. The dramatic changes in the linewidths and level shifts of the two highest and two lowest $H_g$ vibrational modes suggest that the fulleride PP should be anisotropy. In addition the Rb atoms in $Rb_3C_{60}^{3-}$ were expected to have two inequivalent octahedral and tetrahedral positions in the Rb NMR. Experimentally the tetrahedral site is split which also could be explained by an anisotropic PP [34]. It seems that the standard BCS theory offers little hope of providing any explanation. Certainly if the PP suggestion is correct, then it seems likely that the spatial extent of a fulleride PP places it between a "real space" and a momentum space pair. This is not inconsistent with recent theoretical work by Choy et al [35, 36] on high-

$T_c$ systems we discuss later. In addition, recent ARPES dispersion work strongly suggests that pairing without long range order underlies the pseudogap [37].

## 4a. Mean field Theory for Peierls-Fröhlich (Charge Density Wave) Transition. [38, 39]

Treating our two electrons as a 1D free electron gas with a Hamiltonian in second quantized form

$$H = \sum_k \varepsilon_k a_k^\dagger a_k$$

and energy $\varepsilon_k = \hbar^2 k^2 / 2m$ and $a_k^\dagger, a_k$ being the creation and annihilation operators, respectively; spin degrees of freedom are omitted. Following O'Brien [31a], we use an average "collective coordinate" $H_g$ coordinate for all eight vibrations of this symmetry described by Hamiltonian

$$H_{ph} = \sum_q \left\{ \frac{P_q P_{-q}}{2M} + \frac{M\omega_q^2}{2} Q_q Q_{-q} \right\} \qquad (15)$$

where M is the mass, $\omega_q$ are the normal coordinates, and $Q_q$, $P_q$ are the standard normal coordinates and conjugate momenta of the atom motions. This Hamiltonian can be rewritten as

$$H_{ph} = \sum_q \hbar\omega_q \left( b_q^\dagger b_q + \frac{1}{2} \right)$$

using the following

$$Q_q = \left( \frac{\hbar}{2M\omega_q} \right)^{1/2} \left( b_q + b_{-q}^\dagger \right)$$

$$P_q = \left( \frac{\hbar M \omega_q}{2} \right)^{1/2} \left( b_q^\dagger - b_{-q} \right)$$

In this notation the lattice displacement is

$$u(x) = \sum_q \left( \frac{\hbar}{2NM\omega_q} \right)^{1/2} \left( b_q + b_{-q}^\dagger \right) e^{iqx}$$

with $\left( N = \text{lattice-sites} / \text{length} \right)$. Using the "rigid ion approximation" to describe the electron-vibration interaction assumes that the potential V depends only on the distance from the equilibrium lattice position resulting in

$$H_{el-vib} = \sum_{k,k',l} \langle k | V(r-l-u) | k' \rangle a_k^\dagger a_k$$

$$= \sum_{k,k',l} e^{i(k-k')(l+u)} V_{k-k'} a_k^\dagger a_k$$

Here l is the equilibrium atom position, u is the distance from equilibrium, and $V_{k-k'}$ is the Fourier transform of a single atom potential. Approximating for small displacements

$$e^{i(k'-k)u} \approx 1+i(k'-k)u = 1+iN^{-1/2}(k'-k)\sum_q e^{iql}u_q$$

Ignoring the interaction of the electrons with ions in their equilibrium positions, we have

$$H_{el-vib} = iN^{-1/2}\sum_{k,k',l,q} e^{i(k'-k+q)l}(k'-k)u_q V_{k-k'}a_k^\dagger a_k$$

$$= iN^{-1/2}\sum_{k,k'}(k'-k)u_q V_{k-k'}a_k^\dagger a_k$$

Expressing the interaction in second quantization terms,

$$H_{el=vib} = i\sum_{k,k'}\left(\frac{\hbar}{2M\omega_{k-k'}}\right)^{1/2}(k'-k)V_{k-k'}\left(b_{k'-k}^\dagger + b_{k-k'}\right)a_k^\dagger a_{k'}$$

$$= \sum_{k,q} g_q\left(b_{-q}^\dagger + b_q\right)a_{k+q}^\dagger a_k$$

where g, the coupling constant, is

$$g_q = i\left(\frac{\hbar}{2M\omega_q}\right)^{1/2}|q|V_q$$

Putting all three terms together fives what is known as the Fröhlich Hamiltonian

$$H = \sum_k \varepsilon_k a_k^\dagger + \sum_q \hbar\omega b_q^\dagger b_q + \sum_{k,q} g_q a_{k+q}^\dagger a_k\left(b_{-q}^\dagger + b_q\right) \qquad (16)$$

The effect on the normal vibration coordinates for small amplitude displacement is ($\ddot{Q}_q$ is the second derivative with respect to time, and using the equation)

$$\hbar^2 \ddot{Q}_q = -\left[\left[Q_q, H\right], H\right]$$

Since $\left[Q_q, P_{q'}\right] = i\hbar\delta_{q,q'}$, we have

$$\ddot{Q}_q = -\omega_q^2 Q_q - g\left(\frac{2\omega_q}{M\hbar}\right)^{1/2}\rho_q$$

with $\rho_q = \sum_k a_{k+q}^\dagger a_k$ being the $q^{th}$ component of the electron density (g is assumed independent of q). The second term on the RHS is an effective force constant due to combined electron-vibration interaction. The ionic potential $g\left(2M\omega_q/\hbar\right)^{1/2}Q_q$ results in a density fluctuation

$$\rho_q = \chi(q,T)g\left(\frac{2M\omega_q}{\hbar}\right)^{1/2}Q_q$$

and using linear response theory leads to the following equation of motion

$$Q_q = -\left[\omega_q^2 + \frac{2g^2\omega_q}{M\hbar}\chi(q,T)\right]Q_q$$

which produces a renormalized vibration frequency

$$\omega_{ren,q}^2 = \omega_q^2 + \frac{2g^2\omega_g}{M\hbar}$$

For our 1D model $\chi(q,T)$ has its maximum value at $q = 2k_F$, the so-called Kohn anomaly where the reduction (softening) of the vibrational frequency will be most significant. Here

$$\omega_{ren,2k_F}^2 = \omega_{2k_F}^2 - \frac{2g^2 n(\varepsilon_F)\omega_{2k_F}}{\hbar}\ln\left(\frac{1.14\varepsilon_0}{k_B T}\right)$$

As the temperature is reduced, the renormalized vibration frequency goes to zero which defines a transition temperature where a frozen-in distortion occurs. From the equation above a mean field transition temperature for a charge density wave, $T_{CDW}^{MF}$, can be calculated

$$k_B T_{CDW}^{MF} = 1.14\varepsilon_0 e^{-1/\lambda} \qquad (17)$$

with $\lambda$, the electron-vibration coupling constant (dimensionless). This is exactly like the BCS equation; the resulting interaction between CDW and BCS SC is discussed in Section 5.

Below the phase transition we are proposing that the $H_g$ collective renormalized vibration frequency is zero since the lattice distortion is "frozen" as a molecular vibration mode with expectation values $\langle b_{2k_F}\rangle = \langle b_{-2k_F}^\dagger\rangle \neq 0$. Defining an order parameter

$$|\Delta|e^{i\phi} = g\left(\langle b_{2k_F}\rangle + \langle b_{-2k_F}^\dagger\rangle\right) \qquad (18a)$$

The lattice displacement can now be shown to be

$$\langle u(x)\rangle = \left(\frac{\hbar}{2NM\omega_{2k_F}}\right)^{1/2}\left\{i\left(\langle b_{2k_F}\rangle + \langle b_{-2k_F}^\dagger\rangle\right)e^{i2k_F} + cc\right\} \qquad (18b)$$

$$= \left(\frac{\hbar}{2NM\omega_{2k_F}}\right)^{1/2}\frac{2\Delta}{g}\cos(2k_F x + \phi) \qquad (18c)$$

$$= \Delta u \cos(2k_F x + \phi) \qquad (18d)$$

along with

$$\Delta u = \left(\frac{2\hbar}{NM\omega_{2k_F}}\right)^{1/2}\frac{|\Delta|}{g}$$

The Fröhlich Hamiltonian gets modified to

$$H = \sum_k \varepsilon_k a_k^\dagger + \sum_q \hbar\omega b_q^\dagger b_q + \sum_{k,q} g_q a_{k+q}^\dagger a_k \langle b_{-q}^\dagger + b_q\rangle \qquad (19)$$

and since $q = \pm 2k_F$ and $\langle b_{2k_F}\rangle = \langle b_{-2k_F}^\dagger\rangle$

$$H = \sum_k \varepsilon_k a_k^\dagger a_k + 2g\sum_k\left[a_{k+2k_F}^\dagger a_k \langle b_{-2k_F}^\dagger\rangle + a_{k-2k_F}^\dagger a_k \langle b_{-2k_F}\rangle\right] + 2\hbar\omega_{2k_F}\langle b_{2k_F}\rangle^2$$

The order parameter defined earlier (eq 18a) is now

$$H_{el} = \sum_k \left[ \varepsilon_k a_k^\dagger a_k + |\Delta| e^{i\phi} a_{k+2k_F}^\dagger a_k + |\Delta| e^{-i\phi} a_{k-2k_F}^\dagger a_k \right]$$

The standard approximation is to consider only states near the Fermi level. Labeling states near $+k_F$ by index 1 and those near $-k_F$ by index 2, and using the linear dispersion relationship $\varepsilon_k = \hbar v_F (k - k_F)$,

$$H = \sum_k \left[ \varepsilon_k \left( a_{1,k}^\dagger a_{1,k} - a_{2,k}^\dagger a_{2,k} \right) + |\Delta| e^{i\phi} a_{1,k}^\dagger a_{2,k} + |\Delta| e^{-i\phi} a_{2,k}^\dagger a_{1,k} \right]$$

This Hamiltonian can be diagonalized by a canonical transformation with new operators

$$\gamma_{1,k} = U_k a_{1,k} - V_k^\circ a_{2,k} = U_k e^{-i\phi/2} a_{1,k} - V_k e^{i\phi/2} a_{2,k}$$

and

$$\gamma_{2,k} = V_k a_{1,k} + U_k^\circ a_{2,k} = V_k e^{-i\phi/2} a_{1,k} + U_k e^{i\phi/2} a_{2,k}$$

using the constraint $U_k^2 + V_k^2 = 1$. The Hamiltonian becomes

$$H = \sum_k \left[ \varepsilon_k \left( U_k^2 - V_k^2 \right) - 2|\Delta| U_k V_k \right] \left( \gamma_{1,k}^\dagger \gamma_{1,k} - \gamma_{2,k}^\dagger \gamma_{2,k} \right) + \left[ 2\varepsilon_k U_k V_k + |\Delta|\left( U_k^2 - V_k^2 \right) \right] \left( \gamma_{1,k}^\dagger \gamma_{2,k} + \gamma_{2,k}^\dagger \gamma_{1,k} \right)$$

If the coefficients in front of the off-diagonal terms are zero, the Hamiltonian can be diagonalized

$$\left[ 2\varepsilon_k U_k V_k + |\Delta|\left( U_k^2 - V_k^2 \right) \right] = 0 \text{ and } U_k^2 + V_k^2 = 1$$

Using

$$V_k = \cos\left(\frac{\theta_k}{2}\right) \text{ and } U_k = \sin\left(\frac{\theta_k}{2}\right)$$

then

$$\tan \theta_k = -\frac{|\Delta|}{\varepsilon}$$

$$V_k^2 = \frac{1}{2}\left( 1 - \frac{\varepsilon_k}{\left( \varepsilon_k^2 + \Delta^2 \right)^{1/2}} \right) = \frac{1}{2}\left( 1 + \frac{\varepsilon_k}{E_k} \right)$$

and

$$U_k^2 = \frac{1}{2}\left( 1 - \frac{\varepsilon_k}{E_k} \right)$$

where

$$E_k - \varepsilon_F + sign(k - k_F)\left[ \hbar^2 v_F^2 (k - k_F)^2 + \Delta^2 \right]^{1/2}$$

Substituting the expressions for $U_k, V_k$ into the Hamiltonian results in

$$H = \sum_k E_k \left( \gamma_{1,k}^\dagger \gamma_{1,k} + \gamma_{2,k}^\dagger \gamma_{2,k} \right) + \frac{\hbar \omega_{2k_F} \Delta^2}{2g^2} \tag{20}$$

The result is that the linear dispersion $\varepsilon_k - \varepsilon_F = \hbar v_F (k - k_F)$ is no longer valid as a gap is developed in the dispersion. Using a approximation for the density of states

$$N_{CDW}(E)dE = N_e(\varepsilon)d\varepsilon = N_e d\varepsilon$$

$$\frac{N_{CDW}(E)}{N_e} = \frac{d\varepsilon}{dE} = \begin{cases} 0 \\ \dfrac{E}{(E^2-\Delta^2)^{1/2}} \end{cases}$$

The first condition applies if $|E|<\Delta$ while the second when $|E|>\Delta$. The **opening of the gap leads to a lowering of electronic energy**

$$E_{el} = \sum_k (-E_k + v_k k) = n(\varepsilon_F) \int_0^{\varepsilon_F} \left(\varepsilon - (\varepsilon^2 + \Delta^2)^{1/2}\right) d\varepsilon$$

It can be shown that there are two terms leading to this lowering of energy: 1) an electronic term

$$E_{el} = n(\varepsilon_F)\left[-\frac{\Delta^2}{2} - \Delta^2 \log\left(\frac{2\varepsilon_F}{\Delta}\right)\right] + ...$$

and 2) a lattice term

$$E_{latt} = \frac{N}{2} M \omega_{2k_F}^2 \langle u(x)\rangle^2 = \frac{\hbar \omega_{2k_F} \Delta^2}{2g^2} = \frac{\Delta^2 n(\varepsilon_F)}{\lambda}$$

where $\lambda$ is defined earlier. The total energy change then is

$$E = E_{el} + E_{latt} = n(\varepsilon_F)\left[-\frac{\Delta^2}{2} - \Delta^2 \log\left(\frac{2\varepsilon_F}{\Delta}\right) + \frac{\Delta^2}{2\lambda}\right]$$

For $\lambda \ll 1$ and minimizing the total energy gives

$$\Delta = 2\varepsilon_F e^{-1/\lambda} \qquad (21)$$

and a condensation energy of

$$E_{cond} = E_{norm} - E_{CDW} = \frac{n(\varepsilon_F)}{2}\Delta^2 \qquad (22)$$

We propose that the JT stabilization of two electrons ($\varpi$ coordinate in JT Section) and the pseudo 1D nature of the circular CDW are additive. Hence, in 1D the CDW interaction further lowers the potential for the two electrons making the preformed pair binding considerably larger than a "traditional" Cooper pair.

**b. Does the existence of a fulleride preformed pair seem reasonable?**

From an energetic point of view we have illustrated how a JT deformation on the surface of a fulleride can reduce the coulomb potential by using anisotropy to more effectively screen the Coulomb repulsion. An additional JT electronic attraction is the creation of a harmonic well, and a further energetic support is the formation of a charge density wave. Spherical shells of charge appear in a variety of other systems such as multielectron bubbles in helium and metal nanoshells. It seems fruitful to look at the helium example [40]. When a flat surface of helium is charged with electrons above a critical charge density, the surface opens due to instability and a bubble forms with a large number of electrons. A spherical surface is created by the balance of Coulomb repulsion of the electrons countered by the surface tension of helium. The bare electron states on the surface have discrete energies and angular momentum. There are also small amplitude surface waves called spherical ripplons which are quantized.

If this system is treated as a two-dimensional electron gas, the Hamiltonian is

$$H_e^{sphere} = \sum_{L,m,\sigma} \varepsilon_L c^\dagger_{L,m,\sigma} c_{L,m,\sigma} + \sum_{J>0,n} \sum_{L,m,\sigma} \sum_{L',m',\sigma'} v_L c^\dagger_{(L,m)\otimes(J,n),\sigma} c^\dagger_{(L',m')\otimes(J,-n),\sigma'} c_{L',m',\sigma'} c_{L,m,\sigma}$$

with $\sum_{L,m} = \sum_{L=0}^{\infty} \sum_{m=-L}^{L}$ and

$$\varepsilon_L = \frac{\hbar^2}{2m_e} \frac{L(L+1)}{R^2} \text{ and } v_L = \frac{e^2}{2\varepsilon R} \frac{(-1)^L}{2L+1}$$

The theoretical work in these systems suggests that the electron-ripplon coupling might lead to Cooper pairing [41]. A possible viewpoint is that there are limited cases of charged mesoscopic spheres.

## 5. Interaction between SC and CDW

At present quasi one-dimensional systems such as TTF-TCNQ contain CDW, SDW, and superconductivity. Even though the critical temperatures are low, these systems are interesting since in their low dimensions fluctuations of the order parameter suppress ordered phases. The interaction of these three condensed states is not trivial since a SDW breaks time-reversal symmetry, but a CDW does not.

In a quasi one-dimensional system containing superconductivity and CDW studied by Levin et al [42], Bilbro and MacMillan [43], and Balsiero and Falicov [44], the two instabilities are very incompatible. In 2D the phonons can create a CDW metallic state while the electronic state resembles a semiconductor. Thus, the CDW state tends to suppress superconductivity. Machida et al [45] have studied the effect of the CDW state on superconductivity as well as the opposite case; there is a temperature domain where both states coexist.

Of course the advent of HTSC provided more interest in these interactions since some of these were speculated on by Little [46] and Ginsberg [47] as possible new classes of materials which might exhibit "higher" temperature superconductivity. Emery and Kivelson [48] discuss "local electronic structures" or "stripes" with superconductivity appearing when the mentioned stripes become coherent. Also of note is work by Scalapino et al [49] who discuss d wave pairing by exchange of spin and charge density waves. The exact form of their SDW and CDW analysis also fits into the "modified" BFM $(U \neq 0)$. As will be discussed in detail in Section 7, the BCS SC is essentially overpowered by the preformed pair (CDW). It seems that a CDW is one possible physical mechanism whereby the quantum critical point exerts influence on the nature of the surrounding phase diagram.

## 6. Other Pertinent Studies

Geshkenbein, Ioffe, and Larkin [50] discuss the phenomenology of superconductivity resulting from the bose condensation of preformed electron pairs. Despite the difference of the BF model and theirs, several features of their work seem supportive of the BF model. First is that they attribute the phenomena discussed below as attributed to electrons in the corners of the Fermi surface which acquire a gap, followed by coherence peaks at the gap edges? Secondly, converting bosons into fermions on the Fermi surface is allowed. Also, they argue against general bose condensation since it would lead to a huge Hall effect. By assuming the fermions near the corners of the Fermi surface are paired into bosons, charge 2e and no dispersion (a key assumption), we have a representation of this as a circular charge density wave. Using the Hamiltonian

$$H = \sum_{\vec{q}} \varepsilon b_q^\dagger b_q + \sum_{\vec{p},\vec{q}}' V_{\vec{p},\vec{q}} \left( b_{\vec{q}}^\dagger c_{\vec{p}\uparrow} c_{\vec{p}-\vec{q}\downarrow} + hc \right) + \sum_{\vec{p}} \xi_{\vec{p}} c_{\vec{p},\sigma}^\dagger c_{\vec{p},\sigma} \qquad (23)$$

where the sum $\sum'$ excludes the "disc" area. $V_{\vec{p},\vec{q}}$ also has the appropriate symmetry, here d-wave for HTSC, approximated as

$$V_{\vec{p},\vec{q}} = Va^2 \left( d_x^2 - d_y^2 \right) \qquad (24)$$

The comparison with the work of Eliashberg model of strong coupling concerning resonances distributed around the Fermi level is another way of stating the BF model [8]. Bosons become coherent due to exchange of fermions, ultimately the same as the BF model. Finally, they describe a model similar to the Hamiltonian above which describes a phase transition of a system of superconducting grains which contains some similar features to the BF model in the interactions.

**7. The Boson-Fermion Model (BFM) Interpretation of the BCS-BEC Connection.**

The BFM admits a connection between the BCS and BEC regions of the phase diagram (Fig 5) by incorporating a new, strongly attractive interaction between Fermions mediated by the quasi-molecular Boson associated with a Feshbach resonance [4, 8, 15]. Then, applying BCS theory to a degenerate Fermi gas with a strong pairing interaction results in a strong suppression of $T_{BCS}$ caused by fluctuations in the two-particle Cooper channel. In the strong coupling regime there are two "types" of Bosons even above $T_c$, "molecules" ("molecules" in "cold" atom parlance) associated with the resonance and pre-formed pairs, as previously discussed [3, 17], see Fig 1b. As we interpret the model, the pathway from BCS to BEC shows a phase transition due to the resonance in the two body channel. The ground state and single particle excitation gap show the same continuity as previously described [1, 17, 51, 52]

A direct comparison can be made with cold superfluid Fermi gases [1, 3, 4, 8, 15] by reformulating our previous Hamiltonian for clarity (following Ohashi and Griffin [15], and Chen et al [3]) and keeping only the essential terms for the non-resonant and resonant Cooper pair / Boson molecule interaction (see [2], Appendix C for the full Hamiltonian):

$$H = \sum_{\bar{p},\sigma} \varepsilon_{\bar{p}} c^\dagger_{\bar{p}\sigma} c_{\bar{p}\sigma} - U \sum_{\bar{p},\bar{p}'} c^\dagger_{\bar{p}+\bar{q}/2\uparrow} c^\dagger_{-\bar{p}+\bar{q}/2\downarrow} c_{-\bar{p}'+\bar{q}/2\downarrow} c_{\bar{p}'+\bar{q}/2\uparrow} \quad (25)$$

$$+ \sum_q \left(E^0_{\bar{q}} + 2\nu\right) b^\dagger_{\bar{q}} b_{\bar{q}} + g_r \sum_{\bar{p},\bar{q}} [b^\dagger_{\bar{q}} c_{-\bar{p}+\bar{q}/2\downarrow} c_{\bar{p}+\bar{q}/2\uparrow} + h.c.]$$

Here $c_{\bar{p}\sigma}$ and $b_{\bar{q}}$ represent the annihilation operators of a Fermion (Fermi atom) with kinetic energy $\varepsilon_{\bar{p}} = p^2/2m$ and a quasi-molecular Boson with the energy spectrum $E^0_{\bar{q}} + 2\nu = q^2/2M + 2\nu$, respectively. In the second term $-U < 0$ is the BCS theory attractive interaction from non-resonant processes; unfortunately some users of the BFM set U = 0 thereby eliminating the CDW-BCS interaction. The threshold energy of the composite Bose particle energy band is denoted by $2\nu$ in the third term. The last term is the Feshbach resonance (coupling constant $g_r$) that describes how a b-Boson (again, a "molecule" in "cold" atom parlance) can dissociate into two Fermions, or how two Fermions can bind into a b-Boson. Since the b-Boson "molecule" is constructed from a bound state consisting of two Fermions, the boson mass is $M = 2m$ and the conservation of total number of particles N imposes a different number relationship than previously

$$N = \sum_{\bar{p}\sigma} \langle c^\dagger_{\bar{p}\sigma} c_{\bar{p}\sigma} \rangle + 2 \sum_{\bar{q}} \langle b^\dagger_{\bar{q}} b_{\bar{q}} \rangle \equiv N_F + N_B \quad (26)$$

Incorporating this constraint into eq (25) again results in a "grand canonical Hamiltonian", as used previously for variable particle number, since b bosons ("molecules") are formed from fermions, and vice-versa. With this relationship, there is only one chemical potential,

$$H - \mu N = H - \mu N_F - 2\mu N_B$$

and it leads to an energy shift

$$\varepsilon_{\bar{p}} \to \xi_{\bar{p}} \equiv \varepsilon_{\bar{p}} - \mu$$

and

$$\varepsilon_{B\bar{q}} + 2\nu \to \xi_{B\bar{q}} \equiv \varepsilon_{B\bar{q}} + 2\nu - 2\mu$$

From this point we outline a particle-particle vertex and mean-field solutions [30] that provide most of the essential features necessary for the fulleride superconductor phase diagram. We emphasize that these BFM solutions were derived for cold fermion work, but the reformulation exposes what appears to be a surprisingly close connection between the cold Fermion work and fulleride (and presumably high-$T_c$) SC.

The Thouless criterion [53] that describes the instability of the normal phase of Fermions due to an attractive interaction allows formation of pairs. Calculating a four-point vertex function provides an equation for the particle-particle vertex with a solution

$$\Gamma(\bar{q}, i\nu_n) = -\frac{U - g^2 D_0(\bar{q}, i\nu_n)}{1 - [U - g^2 D_0(\bar{q}, i\nu_n)] \Pi(\bar{q}, i\nu_n)} \quad (27)$$

where $D_0^{-1}(\bar{q}, i\nu_n) = i\nu_n - E^0_{\bar{q}} - 2\nu + 2\mu$ is the correlation function of the Cooper pair operator of two Fermions (atoms in cold atom parlance) with total momentum $\bar{q}$ defined by $\hat{\Delta}(\vec{q}) \equiv \sum_{\vec{p}} c^\dagger_{\vec{p}+\vec{q}/2,\uparrow} c^\dagger_{-\vec{p}+\vec{q}/2,\downarrow}$ in the absence of U and $g_r$. The $\Pi(\vec{q}, i\nu_n)$ term is the particle-particle propagator describing Cooper pair fluctuations, needed to describe a non-BCS state in the crossover,

$$\Pi(\vec{q}, i\nu_n) = \sum_{\vec{p}} \frac{1 - f\left(\varepsilon_{\vec{p}+\vec{q}/2} - \mu\right) - f\left(\varepsilon_{\vec{p}-\vec{q}/2} - \mu\right)}{\varepsilon_{\vec{p}+\vec{q}/2} + \varepsilon_{\vec{p}-\vec{q}/2} - 2\mu - i\nu_n} \qquad (28)$$

and $f(\varepsilon)$ is the Fermi distribution. When the particle-particle vertex (eq. 27) develops a pole at $\vec{q} = \nu_n = 0$, a superfluid phase transition occurs which corresponds to the following "gap" equation for $T_c$,

$$1 = \left(U + g_r^2 \frac{1}{2\nu - 2\mu}\right) \sum_{\vec{p}} \frac{\tanh\left(\varepsilon_{\vec{p}} - \mu\right)/2T_c}{2\varepsilon_{\vec{p}} - 2\mu} \qquad (29)$$

In eq (9) $g_r^2/(2\nu - 2\mu)$ is **the additional pairing interaction mediated by a boson which becomes very large when** $2\mu \to 2\nu$. $T_c$ is the temperature at which instability occurs in the normal phase of a degenerate Fermi gas due to formation of bound states with zero center of mass momentum $(\vec{q} = 0)$ and energy $2\mu$.

The chemical potential for this model is determined from eq. (26), ($\mu = \varepsilon_F$ in BCS theory) assuming $\mu$ is temperature independent. However, $\mu$ has a more fundamental deviation when the quasi-molecules with $\vec{q} \neq 0$, pre-formed pairs, and superfluid fluctuations are all present. The "number" equation for Fermions (atoms) $N(\mu, T)$ can be obtained from the thermodynamic potential $\Omega$ as stated previously, $N = -\partial\Omega/\partial\mu$. NSR studied the fluctuations previously [17]; what is new here is the term originating from the Feshbach coupling of the b-Bosons and Fermions in eq. (25). Then,

$$N = N_F^0 + 2N_B^0 - T\sum_{\vec{q}} e^{i\delta\nu_n} \frac{\partial}{\partial\mu} \ln\left[1 - \left(U - g_r^2 D_0(q)\right)\Pi(q)\right] \qquad (30)$$

where $N_F^0 \equiv 2\sum_{\vec{p}} f(\varepsilon_{\vec{p}} - \mu)$ and $N_B^0 = \sum_{\vec{q}} n_B(E_{\vec{q}}^0 + 2\nu - 2\mu)$ with $n_B(E)$ the Bose distribution function. The "gap" and "number" equations (eqs (29) and (30)) can be solved self-consistently. An intuitive interpretation can be obtained by use of the identity $N_B^0 = -T\sum_{\vec{q},\nu_n} D_0(\vec{q}, i\nu_n)$, and rewriting eq. (30) as

$$N = N_F^0 - 2T\sum_{\vec{q}} \tilde{D}(q) - T\sum_{\vec{q}} \frac{\partial}{\partial\mu_F} \ln\left[1 - U_{eff}(q)\right]_{\mu_F \to \mu} \qquad (31)$$

$$\equiv N_F^0 + 2N_B + 2N_C$$

(The artificial separation of the "two types" of Bosons $2N_B$ and $2N_C$ is an attempt to provide further clarity with regard as to their origin.) The second term is a renormalized b-Boson Green function,

$$\frac{1}{\tilde{D}^{-1}(q)} = i\nu_n - E_{\vec{q}}^0 - 2\nu + 2\mu - \Sigma(q)$$

with the self-energy $\Sigma(q) \equiv -g_r^2 \tilde{\Pi}(q)$ and $\tilde{\Pi}(q) \equiv \Pi(q)/[1 - U\Pi(q)]$. This is interpreted as the b-Bosons contribution as affected by the self-energy. The third term $2N_C$ is similar to the fluctuations studied by NSR but now including the pair fluctuations with the effective interaction $U_{eff}(q) = U - g_r^2 D_0(q)$ which now depends on energy as well as momentum.

Transforming the Matsurbara frequency summation into a frequency integration for the renormalized b-Bosons ($N_B$ in eq. (31)), one finds

$$N_B = -\frac{1}{\pi} \sum_{\vec{q}} \int_{-\infty}^{+\infty} dz\, n_B(z) \operatorname{Im} \tilde{D}(\vec{q}, i\nu_n \to z + i\delta)$$

Taking the principle value in the z-integration, we find **that if the b-Boson (molecule) decays into two Fermions in the presence of the Feshbach resonance, it has a finite lifetime** given by the inverse of the imaginary part of the self-energy $-g_r^2 \tilde{\Pi}$ in $\tilde{D}(\vec{q}, z - i\delta)$. But **when the chemical potential becomes negative by lowering the threshold energy $2\nu$, the renormalized b-Bosons do not decay if their energies are smaller than** $E_{\vec{q}}^0 - 2\mu$ since it can be shown from eq. (28) that $im\Pi(\vec{q}, z + i\delta)$ is proportional to the step function $\Theta(z + 2\mu - E_{\vec{q}}^0)$. The energy of a stable molecule $(\omega_{\vec{q}})$ corresponding to the pole of $\tilde{D}$ is given by

$$\omega_{\vec{q}} = \left(E_{\vec{q}}^0 - 2\mu\right) + \left[2\nu - g_r^2 \tilde{\Pi}(\vec{q}, \omega_{\vec{q}})\right] \qquad (32)$$

if $E_{\vec{q}}^0 - 2\mu > 0$. Long-lived Bosons (molecules) appear when the renormalized threshold $2\tilde{\nu} \simeq 2\nu - g_r^2 \tilde{\Pi}(\vec{q}, z)$ becomes negative as decay into two Fermions is forbidden. $N_B$ in eq.(11) consists of two kinds of Bosons, stable ones $N_B^{\gamma=0}$ with infinite lifetime, and quasi-Bosons $N_B^{\gamma>0}$ which can decay into two Fermions. The contribution of the poles describing the stable Bosons gives

$$N_B^{\gamma=0} = \sum_{\vec{q}}^{poles} Z(\vec{q}) n_B(\omega_{\vec{q}}) \quad \text{with} \quad Z(\vec{q})^{-1} = 1 + g_r^2 \frac{\partial \tilde{\Pi}(\vec{q}, \omega_{\vec{q}})}{\partial z}$$

describing the mass renormalization. The third term in eq. (31) describing pair fluctuations can be analyzed similarly. Rewriting the non-resonant s-wave interaction U in terms of the s-wave scattering length $U = 4\pi a N / m$, then $U/\varepsilon_F = 0.85(p_F a)$. For a dilute Fermi gas $p_F a \ll 1$, so $U/\varepsilon_F \ll 1$ is required. Then, the non-resonant attractive interaction U cannot generate pre-formed Cooper pairs in 3D. **However, when $2\mu$ approaches $2\nu$, the interaction $U_{eff}(q)$ mediated by the b-Bosons (molecules) becomes so strong that the pre-formed Cooper pairs appear, as suggested in [17] (see Fig 1).** The energy of these poles is the same as given in eq. (32). This enables us to "divide" $N_C$ into contributions from stable pre-formed Cooper pairs $\left(\equiv N_C^{\gamma=0}\right)$ and scattering states $\left(\equiv N_C^{SC}\right)$. If $g=0$ or $\tilde{\Pi}$ is ignored, there are no pre-formed Cooper pairs and $Z(\vec{q})=1$ in

$$N_C^{\gamma=0} = \frac{g_r^2}{2} \sum_{\vec{q}}^{poles} \frac{\partial \tilde{\Pi}(\vec{q},\omega_{\vec{q}})}{\partial \mu} Z(\vec{q}) n_B(\omega_{\vec{q}})$$

The model Hamiltonian gives an intuitive picture of the BCS to BEC transition. When the threshold $2\nu$ is much larger than $\varepsilon_F$, the Fermi states are dominant. A BCS-like phase transition is found since $\mu \sim \varepsilon_F > 0$ and stable bosons are absent since $\text{Im}\,\tilde{\Pi}(\vec{q},z) \neq 0$ for $z > 0$. The only exception is for $z = 0$, then $\text{Im}\,\tilde{\Pi}(\vec{q},0) = 0$. If $2\tilde{\nu} = 2\mu$ so a stable Boson with $\vec{q}=0$ appears at $\omega_{\vec{q}} = 0$ eq. (32), this condition reduces to the gap equation for $T_c$, eq. (29). This will result in the formation of stable Cooper pair Bosons and a phase transition at the same temperature, namely the BCS theory. In this limit no stable, long-lived Bosons with $\vec{q} \neq 0$ exist above this transition temperature.

In the $\nu \ll 0$ limit, the Fermi states are almost empty and we expect the phase transition to be BEC-like. Stable Bosons can appear even above $T_c$. The phase transition of these stable Bosons occurs when the energy of the Boson $(\vec{q}=0)$ reaches zero as measured from the chemical potential. Again eq. (27) results and the problem is a BEC transition of a non-interacting gas of $N/2$ Bosons with mass $M = 2m$ and no free Fermions. Eq. (27) gives $2\mu = 2\tilde{\nu}$ and a Bose condensate appears in $N_B^{\gamma=0}$ and $N_B^{\gamma \neq 0}$.

The fulleride SC is between the two limits in the previous two paragraphs. It is interesting to note that eq. (29), derived as a condition for a superfluid phase transition via formation of two particle bound states, also describes BEC in a gas of stable b-Bosons and pre-formed Cooper pairs.

## 8. Charge density waves and the Boson-Fermion Model (BFM).

It now seems that most, if not all, of the pieces are in place to offer a more complete model of fulleride and High-T SC. The reformulation of the BF Model in "cold atom" terms made the connections with fulleride SC more apparent, recognizing, of course, that fulleride SC is not a limiting case of BEC. This can be deduced from the phase diagram. If two doped electrons on a fulleride in a singlet state create a preformed pair with bosonic character, then at a low enough temperature it should condense; it doesn't. Therefore, we conclude that the stabilization energy and/or the density of the preformed pair are insufficient or too low, respectively, that the preformed pair is stable. A similar argument applies to the case for four electrons doped on a fulleride, also a singlet. Only in this case the density is higher. An alternative explanation for lack of BE condensation might be topological stabilization of the preformed pair.

Another major difference in the "cold" Fermi-Boson treatment above and the fulleride superconductivity (SC) is that instead of "tuning" $\nu$ using a magnetic field, fulleride superconductivity is tuned by the amount of doped electrons. Starting from the BCS side (Fig 1b) with four doped electrons, when the threshold $2\nu$ is much larger than $\varepsilon_F$, the Fermi states are dominant. A BCS-like phase transition is found since $\mu \sim \varepsilon_F > 0$ and stable bosons are absent ($\operatorname{Im}\tilde{\Pi}(\vec{q},z) \neq 0$ for $z > 0$). The only exception is for $z = 0$ ($\operatorname{Im}\tilde{\Pi}(\vec{q},0) = 0$). If $2\tilde{\nu} = 2\mu$ a stable Boson with $\vec{q} = 0$ appears at $\omega_{\vec{q}} = 0$ in eq. (29), and this condition reduces to the gap equation for $T_c$, eq. (26). This will result in the formation of stable Cooper pair Bosons and a phase transition at the same temperature, namely the BCS theory. In this limit no stable, long-lived Bosons with $\vec{q} \neq 0$ exist above this transition temperature. As doping decreases to, say, somewhere around n = 3.5 electrons, on the way towards the apex in the phase diagram, the unitary point $(\mu = 0)$ is encountered. Electron "molecules" (in the sense of Leggett [54]) with a finite lifetime begin to be formed when $2\nu \leq 2\varepsilon_F$. This is the preformed pair (circular CDW) which performs two functions in BCS-BFM superconductivity. As discussed in section 5, the CDW order parameter competes with the BCS order parameter. So, when the fulleride doping decreases to about 3.5, the preformed pair begins to dominate BCS superconductivity. The second function of the PP is to provide a crucial component of the incipient Feshbach resonance.

Now, the pseuodogap origin seems apparent; it is due to a localized pairing without long range order (LRO), see Fig 1b, as the phases of the wave functions of adjacent fullerides are random. This has been previously recognized [3], but it has also been a puzzle as CDW's compete with SC with the exception of certain organic SC's. The above proposal also suggests why the metal-insulator (M-I) transition that certain doped fullerides span happens; small changes in a fulleride structure can move seemingly closely related fullerides from one side of the transition to the other. The possibility that doping a third electron onto a fulleride can cause an energy change larger than the binding energy of a CDW $E_I$ would suggest that the metal would have three unpaired electrons. This possibility is inconsistent with our superconductivity theory, so we are left with the interesting possibility that the pre formed pair remains intact upon doping with a third electron. The resulting material might be perceived as a "poor conductor" as the possibility exists that the preformed pair remains localized. This material, say $Rb_3C_{60}^{3-}$, might be described as a "boson conductor." This notion receives further support from Choy et al who have managed to remove high energy modes from the Hubbard model and are left with a bosonic component [35, 36]. More specifically, exact integration of the high energy scale results in a charge 2e low energy bosonic mode. There seems to be a strong connection between the BFM and doped Mott-Hubbard-Wigner insulators. As mentioned earlier, recent experimental and theoretical work provide support for pairing with long range order [3, 37, 55]

Continuing our reduction in doping, below doping levels of 2.5 electrons, superconductivity ceases as itinerant "third" electrons are needed to sustain pairing of b-type Bosons. As doping of two electrons is approached, a fulleride molecule maintains a 'pre-formed' Cooper pair (c-type Bosons) as we have described in section 5. The resulting state has Wigner crystal-like correlations and the ordered state is accompanied by the appearance of a preformed pair pseudogap that first began forming when three electrons are doped on a fulleride, Fig 1b.

As mentioned, fulleride SC is not in the BEC limit; if we were dealing with a **true BEC system, in the $\nu \ll 0$ limit**, the Fermi states would be almost empty and we would expect the phase transition to be BEC-like. The phase transition of these stable Bosons occurs when the energy of the Boson $(\vec{q}=0)$ reaches zero as measured from the chemical potential. Again eq. (26) results and the problem is a BEC transition of a non-interacting gas of $N/2$ Bosons with mass $M=2m$ and no free Fermions. Eq. (26) gives $2\mu = 2\tilde{\nu}$ and a Bose condensate appears in $N_B^{\gamma=0}$ and $N_B^{\gamma\neq 0}$. It is interesting to note that eq. (26), derived as a condition for a superfluid phase transition via formation of two particle bound states, also describes BEC in a gas of stable b-Bosons and pre-formed pairs. The end result is that fullerides seem to possess a defined pre-formed CP, a necessary and seemingly illusive point in applying the BF Model to high-$T_c$ systems.

Is there a quantum critical point (QCP) and/or a quantum phase transition (QPT) in the fulleride model? The gap to single particle excitations are given by the minimum of the Bogoliubov quasiparticle energy

$$E_{gap} \equiv \min_{\varepsilon_k \geq 0} \left[ (\varepsilon_k - \mu)^2 + |\Delta_k|^2 \right]^{1/2} \qquad (33)$$

If $\mu > 0$, the minima occur at $\varepsilon_k = \mu$ (BCS) and the energy gap is the gap parameter $\Delta$. As the attraction increases, as some point the chemical potential will go below the bottom of the band where $\varepsilon_k = 0$ and $E_{gap} \neq \Delta$. The gaps to single particle excitation in the s-wave case for both conditions are

$$E_{gap} = \Delta \text{ For } \mu > 0$$

$$\text{and } E_{gap} = \left(\mu^2 + \Delta^2\right)^{1/2} \text{ for } \mu < 0$$

This indicates a weak singularity at $\mu = 0$. At $\mu = 0$ there seems to be a demarcation point between BCS $(\mu > 0)$ and BEC $(\mu < 0)$. As concluded by Engelbrecht et al [49], it does not appear that a singularity is present in the quasiparticle energy. So, the ground state and the single particle excitation are continuous from BCS to BEC.

There is no doubt in the BFM that a Feshbach resonance is present with the result that there is a discontinuity in the two body scattering length (Fig 6). Even though the two body scattering length changes abruptly at the unitary scattering condition, superconductivity still varies smoothly.  **However, there is a fundamental change in the nature of the superconductivity.**  As the unitary limit is traversed going from high doping to low, the BCS Cooper pairs are suppressed as the location of a critical (unitary) point in the fulleride phase diagram is passed.  It seems well established in high-$T_c$ studies that properties are different on one side of the SC "dome" relative to the other [54].  In addition some features of the evidence for a QCP such as quasiparticle lifetime variation have been present [56, 57, 58].  Using the connection we have established with the cold atom work, it seems obvious that at $T = 0$ the Feshbach resonates dominates as it is electronic in nature, whereas other modes usually have the energy depleted.  One of the features of a QCP is presumably its far reaching influence which is clearly present in the Feshbach resonance through the interaction of the preformed pair on BCS SC.  It seems appropriate to suggest that the QCP could result in a QPT [58, 59] from a BCS superconductor to a BFM superconductor.  So, fullerides as a class of compounds could have at least four nearby phases influenced by doping – insulator, metal, BCS and BFM SC, and possibly a fifth, preformed pairs in a metal phase causing a "boson" metal.  Further support for the QPT is obtained in a recent article describing QPT's in the interacting boson model (IBM) in the nuclear case [60].

## 9. Conclusions and Future work

We have illustrated a general BF model for Mott-Hubbard-Wigner insulators and their transition to a number of other (possible five) phases.  The subtlety of these phase transitions seems to offer the possibility of "picking" the phase one wishes to study by doping and other techniques.  The proposal for a quantum phase transition between BCS SC and a near-BEC (or Feshbach resonance) SC is certainly different from the standard QPT discussed in the literature and textbooks.  It is not a ground state phenomenon and as such, it is more subtle, but nonetheless, still has a significant impact on the measured properties.

Avenues for further research are a reassessment of the potential topological nature of the preformed pair (circular CDW).  A second possibility is to continue the connection with accidental spontaneous symmetry breaking; the CDW has the character of a pseudo Goldstone boson (pseudo because it has mass and accidental because the point group symmetry is accidentally a continuous symmetry group within the assumptions of the model).  And certain applying this work more towards high-$T_c$ to establish a similar "CDW" could prove fruitful.  A larger question is are there other superconductors, both existing and not yet discovered, that have a preformed pair?  Lastly, can we quantitatively calculate the CDW condensation energy?  We have not found enough experimental information to complete this task.

Acknowledgement.  RHS wishes to thank Professor Ranninger for clarifying the origin of the BF model.

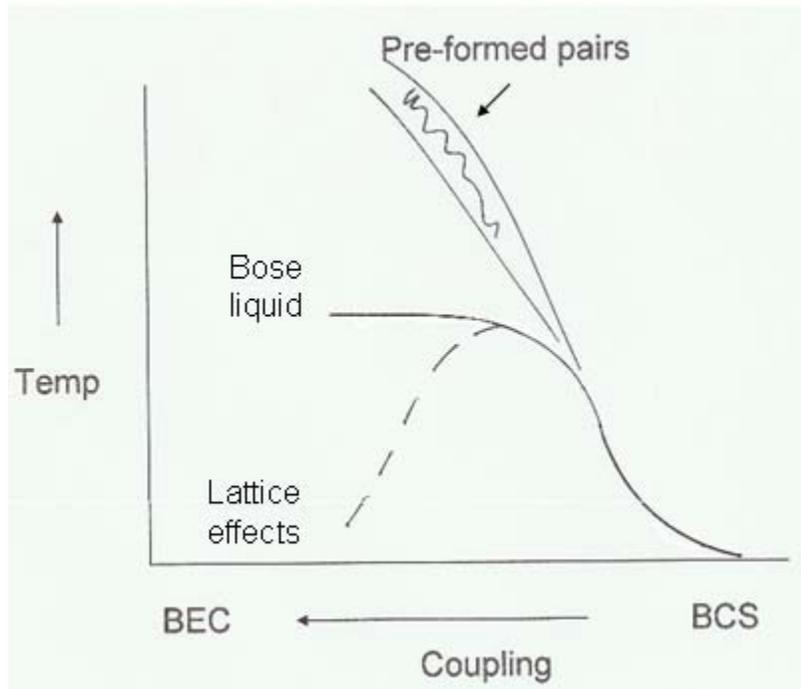

Fig 1a. Qualitative phase diagram for attractive Fermions. The pre-formed pair (broad segment) illustrates a transition region below which pre-formed pairs exist, as in the n = 2 electron doped fulleride (singlet). The full curve is the transition temperature for a continuum model, with the dashed line representing lattice effects. Note the similarities with the fulleride phase diagram in Fig 1b. The coupling increases from right to left.

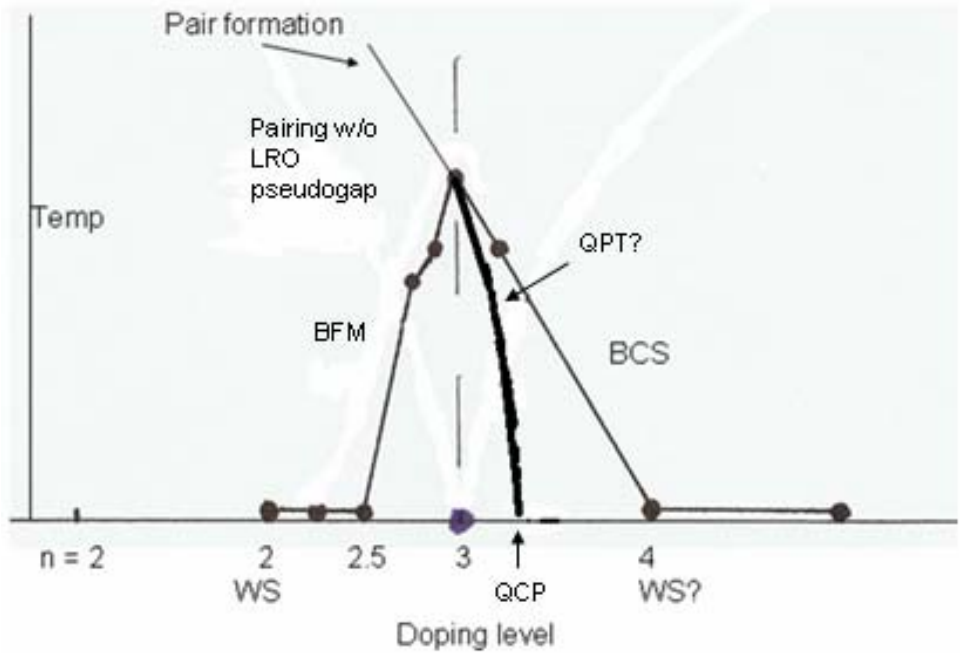

Fig 1b Revised fulleride phase diagram. Pre-formed pairs are the result of the preformed pairing energy being large than the exchange energy for parallel spins in the Jahn-Teller triplet $t_{1u}$ state. The pair is further stabilized by CDW formation (see text). The underdoped BFM region results from Feshbach resonance pairing, while the BCS regime is the result of the non-resonant portion from the pairing term, $U_{eff} = U + \dfrac{g_r^2}{2\mu - 2\nu}$

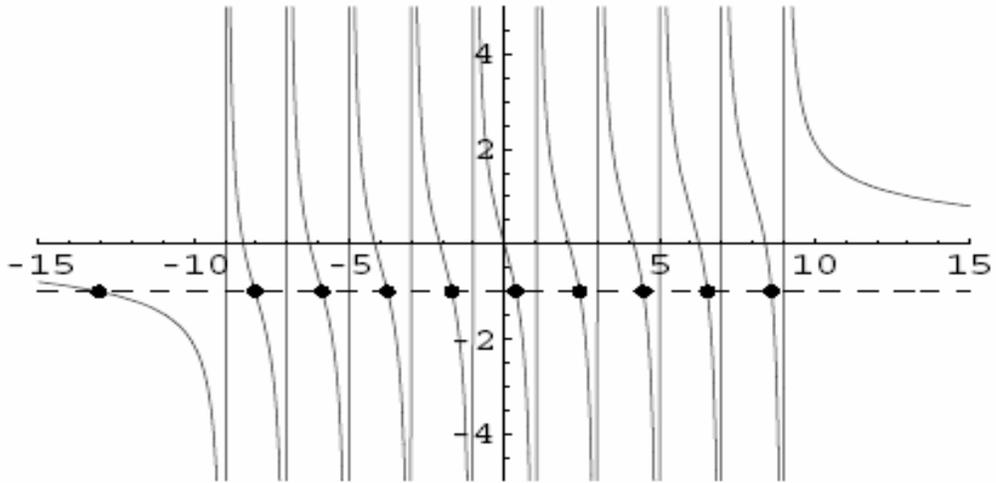

Fig 2. Graphical solution of the Schrödinger equation for the Cooper pair. No matter how weak the attraction, there is always a bound state that breaks away from the continuum. This is a solution for 10 levels.

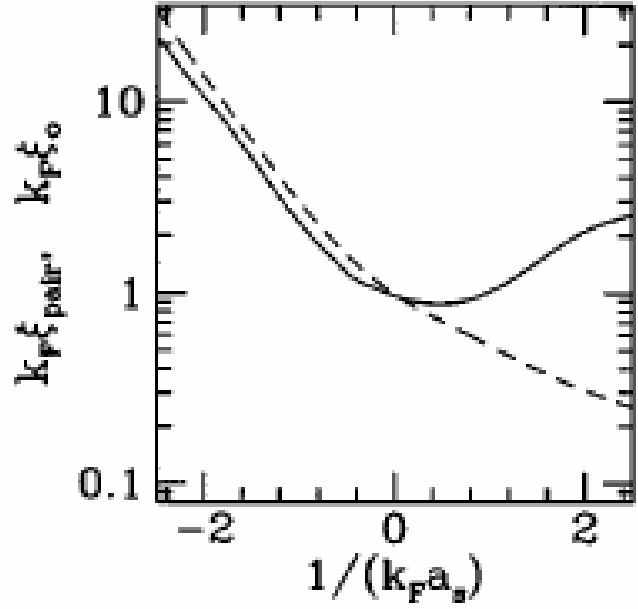

Fig 3 The Ginsburg-Landau (GL) coherence length $\xi_0$ is the solid line (units of $k_F^{-1}$) and the pair size $\xi_{pair}$ is the dashed line plotted as a function of the coupling $1/k_F a_s$. Here the working definition of pair size is $\xi_{pair}^2 = -\langle \psi_k | \nabla_k^2 | \psi_k \rangle / \langle \psi_k | \psi_k \rangle$ and $\psi_k = \Delta_0 / 2E_k$ is the T = 0 pair wave function (after Engelbrecht et al [51]).

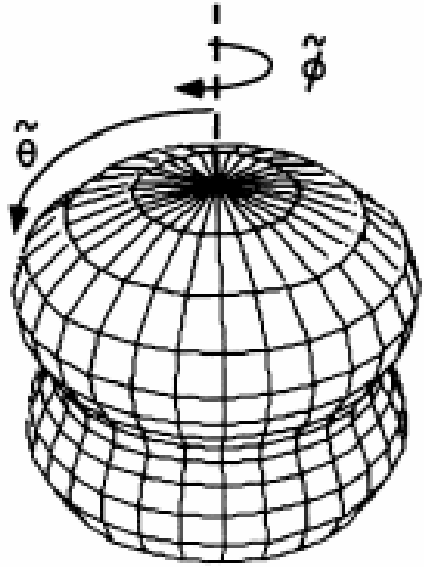

Fig 4. The unimodal Jahn-Teller distortion of a fulleride molecule with two doped electrons in polar coordinates. The equation for the distortion is eq (12) in the text with the equation for the harmonic constraining potential eq (13) and the resulting energies eq (14).

$$\langle u^{JT}(\tilde{\phi},\tilde{\theta})\rangle = \frac{\bar{z}}{2}(3\cos^2\bar{\theta}-1) + \frac{\bar{r}\sqrt{3}}{2}\sin^2\bar{\theta}\cos(2\bar{\phi})$$

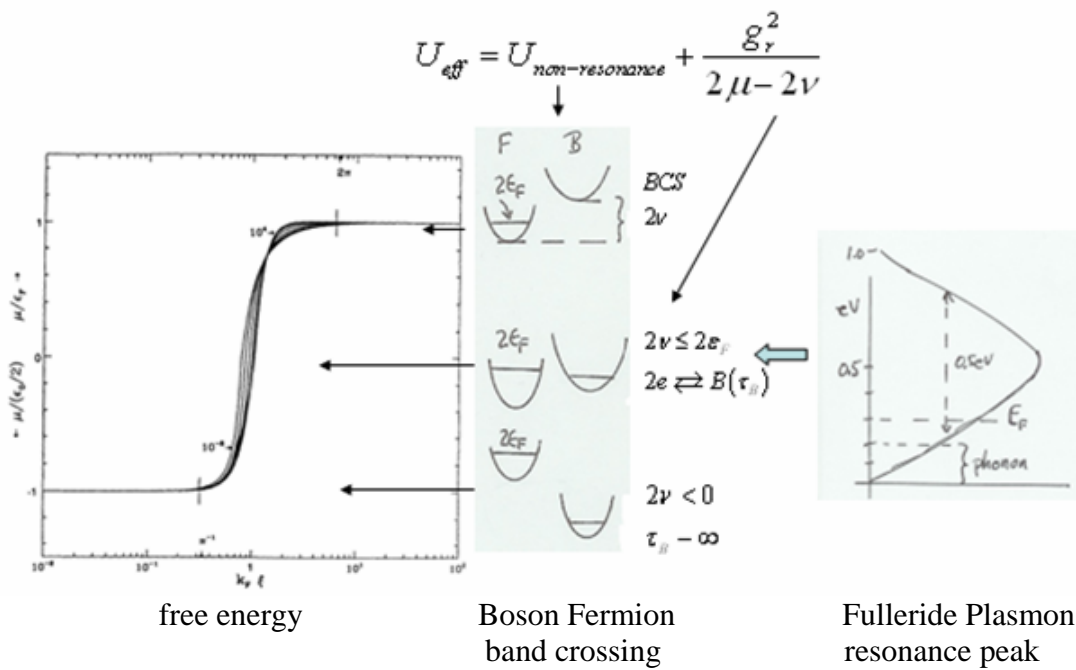

| free energy | Boson Fermion band crossing | Fulleride Plasmon resonance peak |

Figure 5. Correlation of the BCS to BEC transition free energy with the positions of the Fermion (F) and Boson (B) bands and the fulleride resonance peak. The transition is at the unitary limit where the pairing is dominated by the Feshbach resonance. The BEC limit is not reached in either the fullerides or high-$T_c$ superconductivity.

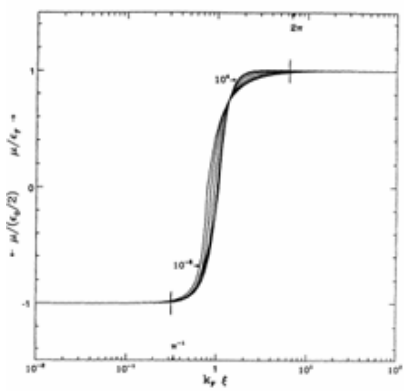 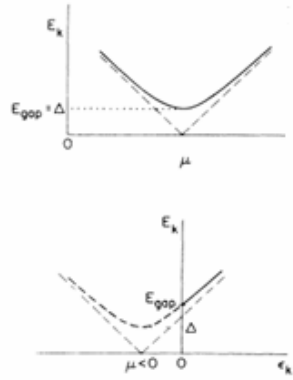 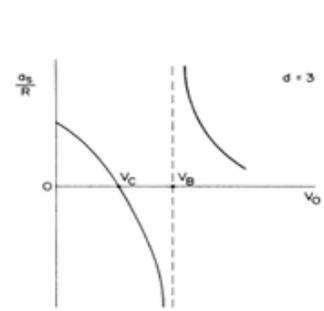

a) free energy      b) quasi-particle energy     c) two body resonance

Fig 6. Illustration of the free energy, quasi-particle energy and two body resonance singularity. In the BCS region ($\mu > 0$) the pairing is relatively weak, and becomes overpowered by the preformed pair as the unitary point (QCP) is approached. Pairing is strong in the BEC region $(\mu < 0)$.